\documentclass[11pt]{article}

\usepackage[includeheadfoot,a4paper,textwidth=170mm,head=6mm,
            headsep=8mm,top=10mm,foot=12mm,bottom=10mm]{geometry}
\usepackage{booktabs} % For formal tables
\usepackage{amssymb,latexsym,amsfonts,amsmath}
\usepackage{graphicx}
\usepackage{enumerate}
\usepackage{multirow}
\usepackage{mathtools}
\usepackage{subfigure}
\newcommand{\pre}[1]{\prescript{#1}{}}
\usepackage[export]{adjustbox}
\usepackage{framed}
\usepackage{tcolorbox}
\usepackage{chngcntr}

\usepackage{wrapfig}
\usepackage{diagbox}

\usepackage[normalem]{ulem}
\usepackage{tikz}
\usetikzlibrary{calc,shapes,arrows}
\usepackage{scalerel}
\usepackage{enumitem}
\usepackage{url}
\usepackage{breakurl}
\usepackage{tikz}
\usepackage{bm}
\usepackage{dsfont}
\usepackage[bb=boondox]{mathalfa}
\usepackage{lipsum}

\newtheorem{theorem}{Theorem}[section]
\newtheorem{lemma}[theorem]{Lemma}

\newtheorem{definition}[theorem]{Definition}

\newtheorem{assumption}[theorem]{Assumption}
\numberwithin{equation}{section}

\newcommand{\II}{\mathsf{id}}

\newcommand{\R}{{\mathbb{R}}}

\newcommand{\TF}{\mathcal{F}}

\newcommand{\N}{{\mathbb{N}}}

\newcommand{\op}{{\mathcal H}}

\newcommand{\SF}{\mathcal{S}}

\newcommand{\n}{\mathcal{N}}
\newcommand{\ca}{\mathcal{C}}

\newcommand{\Let}{:=}

\definecolor{myco}{rgb}{0.2660 0.5740 0.1980}

\newcommand{\Z}{\mathbb{Z}}

\newcommand{\intcc}[1]{\ensuremath{{\left[#1\right]}}}

\title{{\LARGE \bf Symbolic Models for Interconnected Impulsive Systems}\thanks{This work was partly supported by the Google Research Grant, the SERB Start-up Research Grant (RG/2022/001807), the CSR Grants by Siemens and Nokia, the ANR PIA funding: ANR-20-IDEES-0002., and by the German Research Foundation (DFG) as part of Germany’s Excellence Strategy, EXC 2050/1, Project ID 390696704 – “Centre for Tactile Internet with Human-in-the-Loop” (CeTI) of TU Dresden.}}
\author{Sadek Belamfedel Alaoui, Adnane Saoud, Pushpak Jagtap, and Abdalla Swikir 
\thanks{Sadek Belamfedel Alaoui is with the School of Computer Science at Mohammed VI Polytechnical University, Benguerir, Morocco; email: {\tt\small sadek.belamfedel@um6p.ma}. Adnane Saoud is with CentraleSupelec, University Paris-Saclay, Gif-sur-Yvette,  France, and the School of Computer Science at Mohammed VI Polytechnical University, Benguerir, Morocco; email: {\tt\small adnane.soud@centralesupelec.fr}. Pushpak Jagtap is with the Robert Bosch Center for Cyber-Physical Systems, Indian Institute of Science, Bangalore. email: {\tt\small pushpak@iisc.ac.in}.  Abdalla Swikir is with the Chair of Robotics and Systems Intelligence and MIRMI at Technical University Munich (TUM), D-80797 Munich, Germany, the Department of Electrical and Electronic Engineering, Omar Al-Mukhtar University (OMU), QP56 Albaida, Libya and with the Centre for Tactile Internet with Human-in-the-Loop (CeTI), 01062 Dresden, Germany; email: {\tt\small abdalla.swikir@tum.de}.
}}
\onecolumn
\date{}
\begin{document}
\maketitle
\pagestyle{empty}
\thispagestyle{empty}

\begin{abstract}
In this paper, we present a compositional methodology for constructing symbolic models of nonlinear interconnected impulsive systems. Our approach relies on the concept of "alternating simulation function" to establish a relationship between concrete subsystems and their symbolic models. Assuming some small-gain type conditions, we develop an alternating simulation function between the symbolic models of individual subsystems and those of the nonlinear interconnected impulsive systems. 
%This new relation ensures that there is a behavioral mismatch between the compositionality of symbolic models and that of the original interconnected subsystems.
To construct symbolic models of nonlinear impulsive subsystems, we propose an approach that depends on incremental input-to-state stability and forward completeness properties. Finally, we demonstrate the advantages of our framework through a case study.
%\PJ{we should add more about compositionality somewhere!}}\S{Please see again if it s ok?}
%We establish a new methodology for constructing symbolic models of nonlinear interconnected impulsive systems in a compositional manner. The method extends the idea of a simulation function to an approximate alternating simulation function for interconnected systems using the concepts of alternating simulation relations and incremental input-to-state stability. The approximate alternating simulation function ensures that bounded behaviour is preserved between two given subsystems. Our main contribution is a compositional framework for abstracting nonlinear interconnected impulsive systems, which is significantly more efficient in terms of computational time than the monolithic abstraction method. We demonstrate the advantages of our framework through a case study of three interconnected warehouses, which confirms that our approach reduces the computational complexity of symbolic models.
\end{abstract}

\section{Introduction}

The symbolic model (a.k.a abstraction) of dynamical systems involves representing complex systems using finite sets of states, inputs, and transition relations that capture the essential dynamics of the concrete system. The resulting abstract model must be formally included with the concrete system via relations like simulation or alternating simulation \cite{cassandras2009introduction}. This enables model checking and controller design, e.g., through supervisory control and algorithmic game theory. Abstraction-based controller synthesis, commonly used, handles high-level specifications expressed as temporal logic formulae \cite{tabuada2009verification}. However, these approaches depend on state and input space discretization, leading to exponential computational complexity as the concrete system's state space dimension increases. Thus, they face the curse of dimensionality, particularly in high-dimensional systems.

%As applications become more complex, see for example \cite{sadek2023consensus}, it becomes more important to reduce the computational time required to generate their abstractions. This is because the computational resources required for this task can quickly become a bottleneck. Therefore, reducing the computational time of the abstraction process is a challenge to enable the analysis and control of large dynamical systems.

When dealing with complex, interconnected systems, the use of compositional abstraction becomes essential. In this approach, the abstraction process is broken down into smaller subsystem level construction of abstraction, allowing for a more manageable construction of the abstraction of the concrete system.  A significant amount of research has been devoted to developing compositional abstractions for different classes of large-scale interconnected dynamical systems. The results include the construction of compositional abstraction for acyclic interconnected linear \cite{lal2019compositional} , nonlinear \cite{saoud2018compositional}, and discrete-time time-delay \cite{9304184} systems, compositional frameworks based on the notion of an (alternating) simulation function and small-gain type conditions \cite{swikir2019compositional}, compositional frameworks based on dissipativity properties \cite{zamani2017compositional}, compositional abstraction for interconnected switched systems, \cite{8796176,swikir2019compositionalLcss}, and compositional synthesis of abstraction for infinite networks \cite{SWIKIR20201868, SHARIFI2022101173, 9304074, LIU2021101097}, compositional abstraction for interconnected discrete time  systems based on relaxed small-gain conditions \cite{8550268}. A more detailed reference for the compositional framework can be found here \cite{dissertation}. Authors in \cite{saoud2020contract} propose a compositional approach using the concept of assume-guarantee contracts \cite{saoud2021assume}.
%Lavaei et al. \cite{lavaei2017compositional, lavaei2020compositional, lavaei2021compositional} focused on the class of stochastic dynamical systems. On the other hand, these contributions relied on simulation functions and small-gain-type conditions to develop their compositional framework. 
Finally, authors in \cite{saoud2021compositional,sadek2022compositional} proposed compositional abstraction frameworks using the concept of approximate composition.

However, none of the proposed approaches in the literature makes it possible to compositionally construct abstractions for the class of impulsive systems. %Firstly, all the results in the literature relate their abstractions with (approximately alternating) simulation function, where establishing an approximate alternating simulation relation between the abstraction and the concrete system was not addressed.
Indeed, although \cite{swikir2020symbolic} addressed the abstraction of impulsive systems, it focuses on providing a monolithic abstraction of impulsive systems, which can result in a high computational burden when applied to large-scale interconnected systems. Therefore, this paper aims to address this gap in the literature by developing novel results for the compositional abstraction of interconnected impulsive systems. 

This paper establishes a novel compositional scheme for constructing symbolic models of interconnected impulsive systems. In particular, we adapt the notion of alternating approximate simulation functions in \cite{pola2009symbolic} to establish a relation between each subsystem and its symbolic model. Based on some small gain-type conditions, we compositionally construct an overall alternating simulation function as a relation between an interconnection of symbolic models and that of the original interconnected subsystems. Furthermore, under certain stability and forward completeness properties, we present the construction of symbolic models for each subsystem of the original model. 
In our case study, we demonstrate the effectiveness of our approach by comparing the computational efficiency of compositional and monolithic methods for constructing symbolic models of systems while varying the number of interconnected subsystems.

%In our case study, we demonstrate the effectiveness of our approach by constructing the symbolic models of a system consisting of {three} interconnected warehouses using both compositional and monolithic methods. 

%The results show that our approach significantly reduces the computational time required to compute the symbolic model. %In addition, we designed a fixed point controller that effectively regulates the number of items in each warehouse to ensure they remain within the desired range specified for each warehouse.

% These available results in the literature do not address interconnected impulsive systems, which represent room for potential contributions.  

\section{Notations and Preliminaries}\label{1:II}
\paragraph*{\textbf{Notations}}\label{note}
We denote by $\R$, $\Z$, and $\N$ the set of real numbers, integers, and non-negative integers,  respectively.
These symbols are annotated with subscripts to restrict them in
an obvious way, e.g., $\R_{>0}$ denotes the positive real numbers. We denote the closed, open, and half-open intervals in $\R$ by $[a,b]$,
$(a,b)$, $[a,b)$, and $(a,b]$, respectively. For $a,b\in\N$ and $a\leqslant b$, we
use $[a;b]$, $(a;b)$, $[a;b)$, and $(a;b]$ to
denote the corresponding intervals in $\N$.
Given any $a\in\R$, $| a|$ denotes the absolute value of $a$. Given any $u=[u_1;\ldots;u_n]\in\R^{n}$, the infinity norm of $u$ is defined by $\| u\|=\max_{i\in[1;n]}\|u_i\|$.  
%Elements of $\R^n$ are by default regarded as column vectors and we write $\nu\trn$ for the transpose of a vector $\nu\in\R^n$. 
Given a function $\nu: \R_{\ge0} \rightarrow \R^n $, the supremum of $ \nu $ is denoted by $ \| \nu\|_\infty $; we recall that $ \| \nu\|_\infty := \text{sup}_{t\in\R_{\ge0}}\| \nu(t)\|$. Given $\mathbf x:\R_{\geqslant0}\rightarrow\R^{n},\forall t, s\in\R_{\geqslant0}$ with $t\geqslant s$, we define $\mathbf{x}(\pre{-}t)=\lim_{s\rightarrow t}\mathbf{x}(s)$ as the left limit operator. For a given constant $\tau\in\R_{\ge0}$ and a set  $\mathcal{W}:=\{\mathbf x:\R_{\geqslant0}\rightarrow\R^{n}\}$, we denote the restriction of $\mathcal{W}$ to the interval $[0,\tau]$ by $\mathcal{W}|_{[0,\tau]}:=\{\mathbf x:[0,\tau]\rightarrow\R^{n}\}$.
We denote by $\ca(\cdot)$ the cardinality of a given set and by $\varnothing$ the empty set. Given sets $U$ and $S\subset U$, the complement of $S$ with respect to $U$ is defined as $U\backslash S = \{x : x \in U, x \notin S\}$. Given a family of finite or countable sets $S_i, i\in\n\subset\N$, the $j^{th}$ element of the set $S_i$ is denoted by $s_{i_j}$. For any set $S\subseteq\R^n$ of the form $S=\bigcup_{j=1}^MS_j$ for some $M\in \N_{>0}$, where $S_j=\prod_{i=1}^n [c_i^j,d_i^j]\subseteq \R^n$ with $c^j_i<d^j_i$, and non-negative constant $\eta\leqslant\tilde{\eta}$, where $\tilde{\eta}=\min_{j=1,\ldots,M}\eta_{S_j}$ and \mbox{$\eta_{S_j}=\min\{|d_1^j-c_1^j|,\ldots,|d_n^j-c_n^j|\}$}, we define \mbox{$[S]_{\eta}=\{a\in S\,\,|\,\,a_{i}=k_{i}\eta,k_{i}\in\mathbb{Z},i=1,\ldots,n\}$} if $\eta\neq0$, and $[S]_{\eta}=S$ if $\eta=0$. The set $[S]_{\eta}$ will be used as a finite approximation of the set $S$ with precision $\eta\neq0$. Note that $[S]_{\eta}\neq\varnothing$ for any $\eta\leqslant\tilde{\eta}$. 
We use notations $\mathcal{K}$ and $\mathcal{K}_\infty$
to denote different classes of comparison functions, as follows:
$\mathcal{K}=\{\alpha:\mathbb{R}_{\geqslant 0} \rightarrow \mathbb{R}_{\geqslant 0} |$ $ \alpha$ is continuous, strictly increasing, and $\alpha(0)=0\}$; $\mathcal{K}_\infty=\{\alpha \in \mathcal{K} |$ $ \lim\limits_{s \rightarrow \infty} \alpha(s)=\infty\}$.
For $\alpha,\gamma \in \mathcal{K}_{\infty}$ we write $\alpha\le\gamma$ if $\alpha(r)\le\gamma(r)$, $\forall r\in\R_{\geqslant0}$, and, by abuse of notation, $\alpha=c$ if $\alpha(r)=cr$ for all $c,r\geqslant0$. Finally, we denote by $\II$ the identity function over $\R_{\ge0}$, i.e. $\II(r)=r, \forall r\in \R_{\ge0}$.

\subsection{Interconnected Impulsive System} 

\subsubsection{Characterization of Impulsive Subsystems} 

We consider a set of impulsive subsystems indexed by $i\in \n$, where $\n=[1;N]$ and $N\in \mathbb{N}_{\geqslant 1}$. The $ i^{th} $ subsystem can be formally defined by, 
\begin{definition}\label{def:sys1}
A nonlinear impulsive subsystem $\Sigma_i$, $i \in \mathcal{N}$, is defined by the tuple $$\Sigma_i=(\mathbb{R}_{i}^{n_{i}},\mathbb W_i,\mathsf{W}_i,\mathbb U_i,\mathsf{U}_i,f_i,g_i,\mathbb Y_i,h_i,\Omega_i),$$
where 
\begin{itemize}
\item $\R_{i}^{n_i}$ is the state set; 
\item $\mathbb W_i\subseteq\R^{q_i}$ is the internal input set; 
\item $\mathsf{W}_i$ is the set
of all measurable bounded internal input functions $\omega_i:\R_{\geqslant0}\rightarrow \mathbb W_i$;
\item $\mathbb U_i\subseteq\R^{m_i}$ is the external input set;
\item $\mathsf{U}_i$ is the set
of all measurable bounded external input functions $\nu_i:\R_{\geqslant0}\rightarrow \mathbb U_i$;
\item $f_i,g_i: \R^{n_i}\times \mathbb W_i\times \mathbb U_i \rightarrow \R^{n_i} $ are locally Lipschitz functions;
\item $\mathbb Y_i\subseteq\R^{p_i}$ is the output set; 
\item $h{_i} : \mathbb{R}{_i} \rightarrow \mathbb{Y}_{i} $ is the output map;
\item $\Omega_i=\{t_i^k\}_{k\in\N}$ is a set of strictly increasing sequence of impulsive times in $\R_{\ge0}$ comes with $t_i^{k+1}-t_i^{k}\in\{\underline{z}_i\tau_i,\ldots,\overline{z}_i\tau_i\}$ for fixed jump parameters $\tau_i\in\R_{>0}$ and $\underline{z}_i,\overline{z}_i\in \N_{\ge1}$, $\underline{z}_i\le \overline{z}_i$.
\end{itemize}

The non-linear flow and jump dynamics, $ f_{i} $ and $ g_{i} $ are described by differential and difference equations of the form,
\begin{align}\label{eq:2}
\Sigma_i:\begin{cases}
		\mathbf{\dot{x}}_i(t)= f_i(\mathbf{x}_i(t),\omega_i(t),\nu_i(t)), & t\in\R_{\geqslant0}\backslash \Omega_{i},\\
		\mathbf{x}_i(t)=g_i(\mathbf{x}_i(\pre{-}t),\omega_i(\pre{-}t),\nu_i(t)),& t\in \Omega_i,\\
		\mathbf{y}_i(t)= h_i(\mathbf{x}_i(t)),& t\in\R_{\geqslant0},
	\end{cases} 
\end{align}
where $\mathbf{x}_i:\R_{\geqslant0}\rightarrow \R^{n_i}$ and  $\omega_i:\R_{\geqslant0}\rightarrow \mathbb W_i$ are the state and internal input signals, respectively, and assumed to be right-continuous for all $t\in\R_{\geqslant0}$. Function $\nu_i :\R_{\geqslant0}\rightarrow \mathbb U_i$ is the external input signal. We will use $\mathbf{x}_{x_i,\omega_i,\nu_i}(t)$ to denote a point reached at time $t\in \R_{\geqslant0}$ from initial state $x_i$ under input signals $\omega_i\in\mathsf{W}_i$ and $\nu_i\in\mathsf{U}_i$. We denote by $\Sigma_{c_i}$ and $\Sigma_{d_i}$ the continuous and discrete dynamics of subsystem $\Sigma_i$, i.e., $\Sigma_{c_i}:\mathbf{\dot{x}}_i(t)= f_i(\mathbf{x}_i(t),\omega_i(t),\nu_i(t))$, and $\Sigma_{d_i}:\mathbf{x}_i(t)= g_i(\mathbf{x}_i(\pre{-}t),\omega_i(\pre{-}t),\nu_i(t))$. 
\end{definition}
%We say that a hybrid dynamical system is complete if its solution is complete, \cite{tabuada2009verification}. \AS{Do we really need this?}

\subsubsection{Interconnections among Impulsive Subsystems} \label{part2}
We assume that the input-output structure of each impulsive subsystem $\Sigma_i$, $i \in \mathbb{N}$, is general and formally given by,%\vspace{-5pt}
\begin{align}
\label{internalinput}	
 	\omega_i= [\omega_{i1};\dots;\omega_{i(i-1)};\omega_{i(i+1)};\dots;\omega_{iN}], \mathbb W_i\!\!=\!\!\prod_{\substack{j=1,\\j\neq i}}^{N} \!\! \mathbb W_{ij},
  \end{align}
  \begin{align}
\label{output}	
y_i = [y_{i1};\dots;y_{iN}],  \quad \mathbb Y_i=\prod_{j=1}^N  \mathbb Y_{ij},
\end{align}
%\vspace{-5pt}
where $\omega_{ij} \in \mathbb W_{ij}$, $y_{ij} =h_{ij}(x_i)\in\mathbb Y_{ij}$, and output function,
\begin{align}
\label{outputfunction}
&h_i(x_i)\!\! = \!\![h_{i1}(x_i);\dots;h_{iN}(x_i)],
\end{align}
and $ x_{i} $ denotes the state vector of the $ i^{th} $ subsystem. The outputs $y_{i i}$ are considered as external, while $y_{i j}$ with $i \neq$ $j$ are internal and are used to define the connections between the subsystems. In fact, we consider that the dimension of the vector $\omega_{i}$ is equal to that of the vector $y_{i}$. If there is no connection between the subsystems $\Sigma_i$ and $\Sigma_j$, $h_{i j}$ is fixed as zero, i.e. $h_{i j} \equiv 0$. 
\begin{assumption}\label{assumption2.2}
The interconnections are constrained by $\omega_{ij} = y_{ji}$, $\mathbb Y_{ji}\subseteq \mathbb W_{ij}$, $\forall i,j \in  \n , i\neq j$.
%\textcolor{blue}{The interconnections are constrained by $\omega_{ij} - y_{ji}\leqslant \mu_{\kappa} $, $\mathbb Y_{ji}\subseteq \mathbb W_{ij}$, $\forall i,j \in  \n , i\neq j$.}
\end{assumption}

\subsubsection{Interconnected Impulsive Systems}
The formal definition of the interconnected impulsive system is given by,
\begin{definition}
\label{interconnectedsystem} 
Consider $N \in \N_{\geqslant 1}$ impulsive subsystems, $$\Sigma_i=(\R^{n_i},\mathbb W_i,\mathsf{W}_i,\mathbb U_i,\mathsf{U}_i,f_i,g_i,\mathbb Y_i,h_i,\Omega_i)$$ with input-output structure given by \eqref{internalinput}-\eqref{outputfunction}.
The interconnected impulsive system is a tuple $\Sigma = (\mathbb X,\mathbb U,f,\mathcal{G},\Omega)$, denoted by $\mathcal{I}(\Sigma_1,\ldots,\Sigma_N)$ and described by the differential, difference equation of the form,
\begin{align}\label{intdyn}
 \Sigma : \begin{cases}
\dot{\mathbf{x}}(t) = f(\mathbf{x}(t),\nu(t)), & \forall t\in \R_{\geqslant 0}\backslash\Omega \\
\mathbf{x}(t)=\mathcal{G}(\mathbf{x}(\pre{-}t),\nu(t)) & \forall t\in \Omega
\end{cases}
\end{align}
with $x\in \!\mathbb X\!=\!\prod_{i=1}^N \R^{n_i}$, $\nu\! \in \!\mathbb{U}\!=\!\prod_{i=1}^N \mathbb{U}_i$, $\Omega\!=\!\bigcup_{i=1}^N\Omega_{i}$ and 
\begin{align*}
  f(\mathbf{x}(t),\nu(t))= \left[f_{1}(x_1(t),\omega_1(t),v_1(t)), \dots, f_n(x_n(t),\omega_n(t),v_n(t))\right] \\
\mathcal{G}(\mathbf{x}(\pre{-}t),\nu(t))= \left[\beta_{1}(x_1(\pre{-}t),\omega_1(\pre{-}t),v_1(t)), \dots, \beta_n(x_n(\pre{-}t),\omega_n(\pre{-}t),v_n(t))\right]
\end{align*}
where, 
\begin{align*}
\beta_i(x_i(\pre{-}t),\omega_i(\pre{-}t),v_i(t)) =\begin{cases}
    x_i(\pre{-}t) & if \; t \notin \Omega_i \\
    g_i(x_i(\pre{-}t),\omega_i(\pre{-}t),v_i(t)) & if \; t \in \Omega_i
\end{cases}
\end{align*}
\end{definition}

\subsection{Transition systems}

\subsubsection{Transition Subsystems}\label{I}
Now, we will introduce the class of transition subsystems \cite{Tabu}, which will be later interconnected to form an interconnected transition system. Indeed, the concept of transition subsystems permits to model impulsive subsystems and
their symbolic models in a common framework.
	\begin{definition}\label{ts} A transition subsystem is a tuple $T_i=(X_i,X_{0_i},W_i,\mathcal{W}_i,U_i,\mathcal{U}_i,\TF_i,Y_i,\op_i)$, $i\in\mathcal{N}$, consisting of:
		\begin{itemize}
			\item a set of states $X_i$;
		    \item a set of initial states $X_{0_i}\subseteq X_i$;
			\item a set of internal inputs values $W_i$;
			\item a set of internal inputs signals $\mathcal{W}_i\! \! :=\!\! \{\omega_i:\R_{\geq0}\rightarrow W_i\}$;
			\item a set of external inputs values $U_i$;
			\item a set of external inputs signals $\mathcal{U}_i:=\{u_i:\R_{\geq0}\rightarrow U_i\}$;
			\item transition function $\TF_i: X_i\times \mathcal{W}_i\times \mathcal{U}_i \rightrightarrows  X_i$;
			\item an output set $Y_i$;
			\item an output map $\op_i:X_i\rightarrow Y_i$.
		\end{itemize}
		%	\begin{itemize}
		%	\item $X$ is the state set; 
		%	\item $U$ is the input set;
		%	\item $\TF: X\times U \rightrightarrows  X$ is the transition function ;
		%	\item $Y$ is the output set;
		%	\item $\mathcal{H}:X\rightarrow Y$ is the output map.
		%\end{itemize}
	\end{definition}
	The transition $x^+_i\in \TF_i(x_i,\omega_i,u_i)$ means that the system can evolve from state $x_i$ to state $x^+_i$ under the input signals $\omega_i$ and $u_i$. Thus, the transition function defines the dynamics of the transition system. Let $\mathsf{x}_{x_i,\omega_i,u_i}$ denotes an infinite state run of $T_i$ associated with external input signal $u_i$, internal input signal $\omega_i$, and initial state $x_i$. Correspondingly, define $\mathsf{y}_{x_i,\omega_i,u_i}:=\op_i(\mathsf{x}_{x_i,\omega_i,u_i})$ as an infinite output run of $T_i$.
	%Denote the unique value of $u_i\in\mathcal{U}_i$ over $\R$ by $v_{u_i}$.
	Sets $X_i,W_i,U_i$, and $Y_i$ are assumed to be subsets of normed vector spaces with appropriate finite dimensions. 
    If for all $x_i\in  X_i,\omega_i\in \mathcal{W}_i, u_i\in  \mathcal{U}_i$, $\ca(\TF_i(x_i,\omega_i,u_i))\leq1$, we say that $T_i$ is deterministic, and non-deterministic otherwise. Additionally, $T_i$ is called finite if $ X_i,\omega_i, U_i$ are finite sets and infinite otherwise. Furthermore, if for all $x_i\in  X_i$ 
    there exists $\omega_i\in  \mathcal{W}_i$ and $u_i\in  \mathcal{U}_i$ such that $\ca(\TF_i(x_i,\omega_i,u_i))\neq0$ we say that $T_i$ is non-blocking.
%\AS{I don't think the last sentence is true}
%\AS{maybe for all $\omega_i$ there exists $u_i$}

\subsubsection{Interconnections among transition subsystems}
We assume that the input-output structure of each transition subsystem $T_i$, $i \in \mathcal{N}$, is formally defined as the interconnection structure for the impulsive subsystems in part \ref{part2} and is formally defined by,  
\begin{align}	
\label{tinternalinput}	
&\!\!\!\!\!\!\omega_i \!\!=\!\! [\omega_{i1};\dots;\omega_{i(i-1)};\omega_{i(i+1)};\dots;\omega_{iN}], W_i\!\!=\!\!\prod_{\substack{j=1,\\j\neq i}}^{N} \!\! W_{ij},\\
\label{toutput}	
&	y_i = [y_{i1};\dots;y_{iN}], \quad Y_i=\prod_{j=1}^N  Y_{ij},
\end{align}
where $\omega_{ij} \in W_{ij}$, $y_{ij} =h_{ij}(x_i)\in Y_{ij}$, and the output map, 
\begin{align}
\label{toutputfunction}
\!\!\!\!\!\op_i(x_i)\!\! = \!\![\op_{i1}(x_i);\dots;\op_{iN}(x_i)].
\end{align}

%with $\omega_{ij} \in  \omega_{ij}$, $y_{ij} =\op_{ij}(x_i)\in Y_{ij}$.
%	 where $\omega_{ij}$ and $Y_{ij}$ are equipped with $|\cdot|$ and, respectively, $|\cdot|_{Y_{ij}}$ such that  $| y_i|:=\max\limits_{j\in \n}\{| y_{ij}|_{Y_{ij}}\}$, $| \omega_{i}|:=\max\limits_{i\in \n}\{| \omega_{ij}|\}$, for all $y_i\in Y_i$, $\omega_i\in \omega_i$, $y_{ij}\in Y_{ij}$ and $\omega_{ij}\in \omega_{ij}$.
\begin{assumption}\label{assum2.5}
The input-output interconnection variables of transition systems are constrained by,
\begin{equation}\label{eqn:constraint}
\|\omega_{ij} - \op_{ji}(x_j)\|\leqslant \Phi_{ij}, \quad \Phi_{ij} \in \mathbb{R}_{\geq 0}
\end{equation}
\end{assumption}	
% \begin{remark}
% When abstracting subsystems, it is possible that the output of the subsystem does not match the internal input of the direct system. The assumption \ref{assum2.5} allows taking into account this mismatch under a certain tolerance limit $\Phi_{ij}$. 
% \end{remark}

\subsubsection{Composed transition system}
We define the composed transition system by $ \mathcal{I}(T_1,\ldots,T_N) $ and we define it formally by,
\begin{definition}
\label{tinterconnectedsystem} 
Consider $N \in \N_{\geqslant 1}$ transition subsystems $$T_i=(X_i,X_{0_i},W_i,\mathcal{W}_i,U_i,\mathcal{U}_i,\TF_i,Y_i,\op_i)$$ with input-output structure given by \eqref{tinternalinput}-\eqref{outputfunction}. 
The interconnected transition system is a tuple $T=(X,X_{0},U,\TF,Y,\op)$, denoted by $\mathcal{I}(T_1,\ldots,T_N)$, where $X=\prod_{i=1}^N X_i$, $X_0=\prod_{i=1}^N X_{0_i}$, $ U=\prod_{i=1}^N U_i$, $ Y=\prod_{i=1}^N Y_{i}$. Moreover, the transition relation $\TF$ and the output map $\op$ are defined by,
\begin{align}
\label{eqn:transition}
\TF(x,u)&\!\Let\!\{\intcc{x^+_1;\ldots;x^+_N}\,|\, \!x^+_i\in \!\TF_i(x_i,u_i,\omega_i)~ \forall i\!\in\! \n \},\\ \label{eqn:output}
\op(x)&\!\Let\![\op_{11}(x_1);\dots;\op_{NN}(x_N)]
\end{align}
where $x = [x_1;\dots;x_N]\in X$, $u = [u_1;\dots;u_N]\in U$.
%The sets $X$, $U$, and $Y$ equipped with norms $| x|_{X}:=\max\limits_{i\in \n}\{| x_i|_{X_i}\}$, $| u|_{U}:=\max\limits_{i\in \n}\{| u_i|_{\infty}\}$, and $| y|_{Y}:=\max\limits_{i\in \n}\{| y_{ii}|_{Y_{ii}}\}$, respectively, $y = [y_{1};\dots;y_{N}]\in Y$. 

% $ Y_{ji}\subseteq  \omega_{ij}$.
%$|\cdot|=|\cdot|$, $\forall i,j \in  \n , i\neq j$. 
\end{definition}

\subsection{Alternating Simulation Function}

In this section, we recall the so-called notion of $\varepsilon-$ approximate alternating simulation function in \cite{swikir2019compositional}. 

\begin{definition}\label{sf} 
Let $T=(X,X_{0},U,\TF,Y,\op)$ and $\hat T=(\hat{X},\hat{X}_{0},\hat{U},\hat{\TF},\hat{Y},\op)$ with $\hat{Y}\subseteq Y$. A function $ \tilde{\mathcal{S}}:X\times \hat{X} \to \mathbb{R}_{\geqslant0} $ is called an alternating simulation function from $ \hat{T}$ to $\hat T$
if there exist $\tilde{\alpha} \in \mathcal{K}_{\infty}$,  $0<\tilde{\sigma}< 1$, $ \tilde{\rho}_{u} $ $\in \mathcal{K}_{\infty}\cup \{0\} $, and some $\tilde{\varepsilon}\in \mathbb{R}_{\geqslant 0}$ so that the following hold:
\begin{enumerate}
\item For every $ x\in X,\hat{x}\in\hat{X}$, we have,
\begin{align}\label{sf1}
	\tilde{\alpha} (\| \op(x)-\hat{\op}(\hat{x})\|) \!\leqslant\! \tilde{\SF}(x,\hat{x});
\end{align}
\item For every $x\in X,\hat{x}\in\hat{X},\hat u\in\hat{U}$ there exists $u\in U$ such that for every $ x^+\in\TF(x,u)$  there exists $\hat{x}^+\in\hat{\TF}(\hat{x},\hat{u})$ so that,
\begin{align}\label{sf2}
	\tilde{\SF}&(x^+,\hat{x}^+)\leqslant \max\{\tilde{\sigma} \tilde{\SF}(x,\hat{x}),\tilde{\rho}_u(\|\hat{u}\|_\infty ),\tilde{\varepsilon}\};
\end{align}
% \item For every $x\in X,\hat{x}\in\hat{X},u\in U$ there exists $\hat{u}\in \hat{U}$ such that for every $ \hat x^+\in\hat \TF(\hat x,\hat u)$  there exists ${x}^+\in{\TF}({x},{u})$ so that, 
% \begin{align}\label{sf3}
% 	\tilde{\SF}&(x^+,\hat{x}^+)\leqslant \max\{\tilde{\sigma} \tilde{\SF}(x,\hat{x}),\tilde{\rho}_u(\|{u}\|_\infty ),\tilde{\varepsilon}\}.
% \end{align}
\end{enumerate}
\end{definition}
It was shown in \cite{swikir2019compositional} that the existence of an approximate alternating simulation function implies the existence of an approximate alternating relation from $T$ to $\hat{T}$. This relation guarantees that for any output behavior of $T$ there exists one of $\hat T$ such that the distance between these two outputs is uniformly bounded by $\hat{\varepsilon}=\tilde{\alpha}^{-1}(\max\{\tilde{\rho}_{u}(r),\tilde{\varepsilon}\})$. For local abstraction, the notion of $\varepsilon$-approximately alternating simulation function from  $T_i$ to $\hat T_i$, $ \forall i \in \mathcal{N} $, is formally defined by,

% From the perspective of controller design, the concept of an approximate alternating simulation relation is better than approximate simulation. Compared to the approximate simulation relation in \cite{POLA20082508}, if there is no controller for $\hat{T}$, one can conclude that there is no controller for $T$ either.

% \begin{remark}
% Since inputs are bounded in all practical applications, the requirement that control inputs be bounded is not restrictive at all. Moreover, under certain properties of impulsive systems (see Section \ref{1:IV}), one can choose the function $\tilde{\rho}_{u}$ in \eqref{er} to be identically zero, which removes the dependence on the size of the control inputs in Proposition \ref{error}.
% \hfill$\diamond$
% \end{remark}

%	\AS{it needs to be modified, since we didn't defined the alternating simulation} We say that $\hat T$ is an abstraction of $T$ if there exists an alternating simulation function from $\hat T$ to $T$. In addition, if $\hat T$ is finite ($\hat X$ and $\hat U$ are finite sets), we say that $\hat T$ is a symbolic model of $T$.

\begin{definition}\label{def:SFD1}
Let $T_i=(X_i,X_{0_i},W_i,U_i,\TF_i,Y_i,\op_i)$  and $\hat T_i=(\hat X_i,\hat X_{0_i},\hat W_i,\hat U_i,\hat \TF_i,\hat Y_i,\hat \op_i)$ be transition subsystems with $\hat Y_i\subseteq Y_i$, $\hat \omega_i\subseteq W_i$. A function $ \mathcal{S}_i:X_i\times \hat X_i \to \mathbb{R}_{\geqslant0} $ is called a local alternating simulation function from $\hat T_i$ to $T_i$
if there exist $\alpha_i, \rho_{\omega_i}\in \mathcal{K}_{\infty}$,  $0<\sigma_i< 1$, $ \rho_{u_i} \in \mathcal{K}_{\infty}\cup \{0\} $, and some $\varepsilon_i\in \mathbb{R}_{\geqslant 0}$ so that the following hold:
\begin{enumerate}
\item For every $ x_i\in X_i,\hat{x}_i\in\hat{X}_i$, we have, 
\begin{align}\label{lsf1}
	\alpha_i (\|\op_i(x_i)-\hat{\op}_i(\hat{x}_i)\| ) \!\leqslant\! \SF_i(x_i,\hat{x}_i);
\end{align}
\item For every $x_i \!\in\!  X_i, \hat x_i \!\in\! \hat{X}_i,\hat{u}_i \!\in\!\hat{U}_i$ there exists $ u_i \!\in\! U_i$ such that for every $ \omega_i\in W_i, \hat \omega_i\in\hat{W}_i, x^+_i \!\in\! \TF_i(x_i,\omega_i,u_i)$ there exists $\hat{x}^+_i \!\in\! \hat{\TF}_i(\hat{x}_i,\hat{\omega}_i,\hat{u}_i)$ so that,
\begin{align}\label{lsf2}
	\SF_i(x^+_i,\hat{x}^+_i)\leqslant &		 \bar{\sigma}_i \SF_i(x_i,\hat{x}_i)+ \bar\rho_{\omega_i} (\| \omega_i\!-\!\hat \omega_i\|)\nonumber\\
	&+\bar\rho_u(\|\hat{u}_i\|_{\infty} )+\bar\varepsilon_i.
\end{align}
% \item For every $x_i\in  X_i, \hat x_i\in {\hat{X}}_i, u_i\in{U}_i$ there exists $ \hat u_i\in \hat{U}_i$ such that for every $ \omega_i\in \omega_i, \hat \omega_i\in\hat{\omega}_i $, $\hat{x}^+_i \in \hat{\TF}_i(\hat{x}_i,\hat{\omega}_i,\hat{u}_i)$ there exists $x^+_i \in \TF_i(x_i,\omega_i,u_i)$ so that,
% \begin{align}\label{lsf3}
% 	\SF_i(x^+_i,\hat{x}^+_i)\leqslant& \bar\sigma_i \SF_i(x_i,\hat{x}_i)+ \bar\rho_{\omega_i} (\| \omega_i\!-\!\hat \omega_i\|)\nonumber\\
% 	&+\bar\rho_u(\|{u}_i\|_{\infty} )+\bar\varepsilon_i.
% \end{align}
\end{enumerate}
\end{definition}
%	$\hat{T}_i$ is called an abstraction of $T_i$ if there exists a local alternating simulation function from $\hat{T}_i$ to $T_i$. Additionally, if $\hat{T}_i$ is finite ($\hat{X}_i$, $\hat{\omega}_i$ and $\hat{U}_i$ are finite sets), $\hat{T}_i$ is called a symbolic model of $T_i$.

The goal is to construct alternating simulation functions for the compound transition systems $T=\mathcal{I}(T_1,\ldots,T_N)$ and $\hat T=\mathcal{I}(\hat T_1,\ldots,\hat T_N)$ from the local alternating simulation functions of the subsystems. 
To achieve this goal, the following lemmas are recalled.

\begin{lemma} \cite[Theorem $1$]{arxiv}\label{sfm} 
Let $ \mathcal{S}_i:X_i\times \hat X_i \to \mathbb{R}_{\geqslant0} $ be a local alternating simulation function from $\hat T_i$ to $T_i$ then, 
% \begin{enumerate}
for every $x_i \!\in \! X_i, \hat x_i \!\in \! {\hat{X}}_i,\hat u_i \!\in\hat{U}_i$ there exists $ u_i \!\in {U}_i$ such that for every $ \omega_i \!\in \! W_i, \hat \omega_i \!\in\hat{W}_i, x^+_i  \!\in \! \TF_i(x_i,\omega_i,u_i)$ there exists $\hat{x}^+_i  \!\in \! \hat{\TF}_i(\hat{x}_i,\hat{\omega}_i,\hat{u}_i)$ so that, 
\begin{align}\label{lsmf2}
\SF_i(x^+_i,\hat{x}^+_i) \leqslant&\max \left\lbrace \sigma_i \SF_i(x_i,\hat{x}_i), \rho_{\omega_i} (\| \omega_i\!-\!\hat \omega_i\|), \right. \nonumber\\
&\left.\rho_{u_i}(\|\hat{u}_i\|_{\infty} ),\varepsilon_i\right\rbrace;
\end{align}
% \item For every $x_i \!\in  X_i, \hat x_i \!\in {\hat{X}}_i, u_i \!\in{U}_i$ there exists $ \hat u_i \!\in \hat{U}_i$ such that for every $ \omega_i \!\in \! \omega_i, \hat \omega_i \!\in\hat{\omega}_i, $ $\hat{x}^+_i  \!\in \hat{\TF}_i(\hat{x}_i,\hat{\omega}_i,\hat{u}_i)$ there exists $x^+_i  \!\in \TF_i(x_i,\omega_i,u_i)$ so that,
% \begin{align}\label{lsmf3}
% 	\SF_i(x^+_i,\hat{x}^+_i)\leqslant& \max\left\lbrace\sigma_i \SF_i(x_i,\hat{x}_i), \rho_{\omega_i} (\| \omega_i(0)\!-\!\hat \omega_i(0)\|),\right.\nonumber\\
% 	&\left.\rho_{u_i}(\|{u}_i\|_{\infty} ),\varepsilon_i\right\rbrace.
% \end{align}
% \end{enumerate}
where ${\sigma_i}=1-(1-\psi)(1-{\bar{\sigma}_i})$, $\rho_{\omega_i}=\frac{1}{(1-{\bar{\sigma}})\psi}{\bar{\rho}_{\omega_i}}$, $ \rho_{u_i}=\frac{1}{(1-{\bar{\sigma}})\psi}{\bar{\rho}_{u_i}}$, and ${\varepsilon_i}=\frac{\bar{\varepsilon}}{(1-{\bar{\sigma}_i})\psi}$, for an arbitrarily chosen positive constant $\psi<1$, and $\bar{\sigma},\bar{\varepsilon},\bar{\rho}_w,\bar{\rho}_u$ are constants and function appearing in Definition \ref{def:SFD1}.
\end{lemma}

% \begin{lemma}\label{lem2}\cite{Kellett2014}
% Consider $\alpha  \in \mathcal{K}$ and $\chi \in \mathcal{K}_{\infty}$, where $(\chi-\mathcal{I}_d)\in \mathcal{K}_{\infty}$. Then for any $a,b \in \R_{\geqslant0}$ 
% \begin{align*}%\label{lem1}
% \alpha(a+b)\leqslant\alpha\circ\chi(a)+\alpha\circ\chi\circ(\chi-\mathcal{I}_d)^{-1}(b).
% \end{align*}
% \end{lemma}

% \begin{lemma} \label{lem1}\cite{swikir2019compositional}
% For any $a, b \in \mathbb{R}_{>0}$, the following holds
% \begin{align*}
% a+b \leq \max \left\{\left(\mathcal{I}_d+\lambda\right)(a),\left(\mathcal{I}_d+\lambda^{-1}\right)(b)\right\}
% \end{align*}
% for any $\lambda \in \mathcal{K}_{\infty}$.    
% \end{lemma}

\section{Compositionality Result}
The goal of this section is to provide a method for the compositional construction of an alternating simulation function for the interconnected transition system $T=\mathcal{I}(T_1,\ldots,T_N)$ to $\hat T=\mathcal{I}(\hat T_1,\ldots,\hat T_N)$ as defined in Definition \ref{tinterconnectedsystem}. For the functions $\sigma_i$, $\alpha_i$, and $\rho_{wi}$ associated with $\SF_i$, $i\in \mathcal{N}$, given in Lemma \ref{sfm}, we define $\forall i,j \in \mathcal{N}$,
\begin{align}\label{gammad}
\gamma_{ij}\Let\begin{cases}
\sigma_{i}&  \text{if} \quad i,j \in \n | i=j,\\
\rho_{\omega_i}\circ\alpha_{j}^{-1}& \text{if} \quad i,j \in \n | i\neq j,
\end{cases}
\end{align}
and we set $\gamma_{ij}$ equal to zero if there is no connection from $T_j$ to $T_i$, i.e., $\omega_{ij}=0$.

To establish the compositionality results of the paper, we make the following scaled small-gain assumption.
\begin{assumption}\label{sg}
Assume that functions $\gamma_{ij}$ defined in \eqref{gammad} satisfy,
\begin{align}\label{SGC}
\gamma_{i_1i_2}\circ\gamma_{i_2i_3}\circ\cdots\circ\gamma_{i_{r-1}i_r}\circ\gamma_{i_ri_1}<\II,
\end{align}
$\forall(i_1,\ldots,i_r)\in\{1,\ldots,N\}^r$, where $r\in \{1,\ldots,N\}$.\\
\end{assumption}
% \begin{remark}
% Note that from Theorem 5.2 in \cite{090746483}, the scaled small-gain condition \eqref{SGC} implies the existence of $\psi_i \in \mathcal{K}_{\infty}$ $\forall i\in  \n $, satisfying, 
% \begin{align}\label{gam}
% &\max\limits_{j\in  \n }\{\psi^{-1}_i\circ\gamma_{ij}\circ\psi_j\}<\II.
% \end{align} 
% \end{remark}
% \begin{remark}
% To compute the $\mathcal{K}_{\infty}$ functions $\psi_i, i\in \n $ it is possible to consider three cases:
% \begin{itemize}
% \item[(i)] If $N\in \{2,3\}$, i.e., the system is made of two or three subsystems, then one can use the simple construction techniques provided by \cite{JIANG} and \cite[Section 9]{090746483};
% \item[(ii)] If $N\in \N \setminus \{2,3\}$, functions $\psi_i, i\in \n $, can be constructed numerically using the algorithm proposed by \cite{Eaves} and the technique provided by \cite[Proposition 8.8]{090746483}, see \cite[Chapter 4]{Rufferp};
% \item[(iii)] If $\gamma_{ij}<\II$, $\forall~ i,j\in  \n $, then the functions $\psi_i, i\in \n $, can be chosen as identity. Moreover, inequality \eqref{OF} reduces to $\SF(x,\hat{x})\Let\max\limits_{i\in \n }\{ \SF_i(x_{i},\hat{x}_{i})\} $. 
% \end{itemize}
% \end{remark}
The next theorem provides a compositional approach to construct an alternating simulation function from $\hat T=(\hat T_1,\ldots,\hat T_N)$ to $T=(T_1,\ldots,T_N)$ via local alternating simulation functions from $\hat{T}_i$ to $T_i$, $ i \in \mathcal{N} $. 

\begin{theorem}\label{thm:3}
Consider the interconnected transition system $T=\mathcal{I}(T_1,\ldots,T_N)$. Assume that each $T_i$ and its abstraction $\hat{T}_i$ admit a local alternating simulation function $\SF_i$ as in Lemma \ref{sfm}.
Suppose Assumption \ref{sg} holds.
Then, function $\tilde{\SF}:X\times \hat{X}\rightarrow \R_{\ge0}$ defined as,
\begin{align}\label{OF}
\tilde{\SF}&(x,\hat{x})\Let\max\limits_{i\in \n }\{ \psi^{-1}_{i} (\SF_i(x_{i},\hat{x}_{i})) \} 
\end{align}
is an alternating simulation function from $T=\mathcal{I}(T_1,\ldots,T_N)$ to  $\hat T=\mathcal{I}(\hat T_1,\ldots,\hat T_N)$. 
\end{theorem}

\bigskip

% \begin{remark}\label{nosgc}
% If there exist at least one pair $ a,b \in \n $ satisfying $ \gamma_{ab}<\II $ and for any $ i,j\in \n \backslash (a,b) $ $\gamma_{ij}=\II, $ then, assumption \eqref{SGC} still holds and we can consider, $\psi_{i}=\II$ for all $i\in \n $. Moreover, inequality \eqref{OF} reduces to $\SF(x,\hat{x})\Let\max\limits_{i\in \n }\{ \SF_i(x_{i},\hat{x}_{i})\} $.
% \hfill$\diamond$ 
% \end{remark}

% \S{old remark:\\
% \begin{remark}\label{nosgc}
% 			If $\gamma_{ij}\leq\II$ for any $i,j\in[1;N]$, consequently, $\psi_{i}=\II$ for all $i\in[1;N]$, the small-gain condition \eqref{SGC} is satisfied automatically. Moreover, inequality \eqref{OF} reduces to $\SF(x,\hat{x})\Let\max\limits_{i\in[1;N]}\{ \SF_i(x_{i},\hat{x}_{i})\} $.
% 			\hfill$\diamond$ 
% 		\end{remark}
% }

\section{Construction of Symbolic Models}\label{1:IV}

In the previous section, we showed how to construct an abstraction of a system from the abstractions of its subsystems. In this section, our focus is on constructing a symbolic model for an impulsive subsystem using an approximate alternating simulation. To ease readability, in the sequel, the index $i\in \n $ is omitted.

Consider an impulsive subsystem $\Sigma=(\R^{n},\mathbb W,\mathsf{W},\mathbb U,\mathsf{U}_{\tau},f,g,\mathbb Y,h,\Omega)$, as defined in Definition \ref{def:sys1}. We restrict our attention to sampled-data impulsive systems, where the input curves belong to $\mathsf{U}_{\tau}$ containing only curves of constant duration $\tau$, i.e.,
\begin{align}\label{input}
\mathsf{U}_{\tau}=\{\nu:\R_{\ge0}\rightarrow \mathbb U| \nu(t)&=\nu((k-1)\tau),\\\notag &t\in [(k-1)\tau,k\tau),k\in\N_{\geqslant1}\}.
\end{align}
Moreover, we assume that there exist constant $\varphi$ such that for all $\omega\in \mathsf{W}$ the following holds,
\begin{align}\label{gb}
\|\omega(t)-\omega((k-1)\tau)\|\leqslant\varphi,\forall t\in [(k-1)\tau,k\tau),k\in\N_{\geqslant1}.
\end{align}

We also have the following Lipschitz continuity assumption on the output map $h$.

\begin{assumption}\label{h} 
There exist positive constant $L$, such that the output maps $h$ satisfy the following Lipschitz assumption is satisfied, 
\begin{align}\label{lc} 
\| h(x)\!-\!h(y)\|\!\leqslant\! L\| x \!-\! y\|~ \forall x,y\in \R^{n}.
\end{align}
\end{assumption}

% \begin{remark}
% Since the internal input $\omega$ results from the concatenation of the neighbors' outputs. It suffices to show that $\forall x \in X $ the  term $\|y((k+1)\tau)-y(k\tau)\| $ is bounded. 

% Note that the physical variables are restricted to a compact set, thus, the states and inputs have bounded quantities, which implies from the continuity of the maps $f$ and $g$ that there exists a positive constant $C^{te}$ such that $ \forall x \in X $ we have, $ \|x((k+1)\tau)-x(k\tau)\|
% \leqslant C^{te} \tau $, which gives,
% \begin{align*}
% \|y((k+1)\tau)-y(k\tau)\| \leqslant& \|h(x((k+1)\tau))-h(x(k\tau))\| \\
% \leqslant& L\|x((k+1)\tau)-x(k\tau)\|\\
% \leqslant& L C^{te} \tau = \varphi
% \end{align*}
% hence, condition \eqref{gb} is not restrictive.
% \end{remark}

%		For later use, define $\mathcal{W}_{\tau}$ as
%		 \begin{align}\label{ininput}
%		\mathcal{W}_{\tau}=\{\omega:\R_{\ge0}\rightarrow W| \omega(t)&=\omega((k-1)\tau),\\\notag &t\in [(k-1)\tau,k\tau),k\in\N_{\geqslant1}\}.
%		\end{align}
%			For later use, define $\mathcal{A}(\mathcal{W}_{i})$ as
%		\begin{align}\label{ininput}
%	\mathcal{A}(\mathcal{W}_{i})=\{\overline{\omega}:\R_{\ge0}&\rightarrow W| \overline{\omega}(t)=\overline{\omega}((k-1)\tau)=\omega(0),\\\notag &t\in [(k-1)\tau,k\tau),k\in\N_{\geqslant1},\omega\in\mathcal{W}_{i}\}.
%		\end{align}
Next, we define sampled-data impulsive systems as transition subsystems. Such transition subsystems would be the bridge that relates impulsive systems to their symbolic models.
\begin{definition}\label{tsm} Given an impulsive system $\Sigma=(\R^{n},\mathbb W,\mathsf{W},\mathbb U,\mathsf{U}_{\tau},f,g,\mathbb Y,h,\Omega)$, we define the associated transition system $T_{\tau}(\Sigma)=(X,X_{0}, W,\mathcal{W},U,\mathcal{U},\TF,Y,\op)$ 
where:
\begin{itemize}
\item  $X=\R^{n} \times \{0,\ldots,\overline{z}\}$;
\item  $X_{0}=\R^{n}\times \{0\}$; 
\item $U=\mathbb U$;
\item $\mathcal{U}=\mathsf{U}_{\tau}$;
\item $W=\mathbb W$;
\item $\mathcal{W}=\mathsf{W}$;
\item  $(x^+,c^+)\in \TF((x,c),\omega,u)$ if and only if one of the following scenarios hold:
\begin{itemize} 
\item Flow scenario: $0\leq c\leq \overline{z}-1$, $x^+= \mathbf x_{x,\omega,u}(\pre{-}\tau)$, and $c^+=c+1$;
\item Jump scenario: $\underline{z}\leq c\leq \overline{z}$,  $x^+= g(x,\omega(0),u(0))$, and $c^+=0$; 
\end{itemize}
\item $Y=\mathbb Y$;
\item $\mathcal{H}:X\rightarrow Y$, defined as $\mathcal{H}(x,c)=h(x)$.
\end{itemize}
\end{definition}
For later use, define $\mathcal{W}_{\tau}$ as,
\begin{align}\label{ininput}
\mathcal{W}_{\tau}=\{\omega:\R_{\ge0}\rightarrow W| \omega(t)&=\omega((k-1)\tau),\\\notag &t\in [(k-1)\tau,k\tau),k\in\N_{\geqslant1}\}.
\end{align}
In order to construct a symbolic model for $T_{\tau}(\Sigma)$, we introduce the following assumptions and lemmas. 

\begin{assumption}\label{likeiss}
Consider impulsive system $\Sigma=(\R^{n},\mathbb W,\mathsf{W},\mathbb U,\mathsf{U}_{\tau},f,g,\mathbb Y,h,\Omega)$. Assume that there exist a locally Lipschitz function $ V:\R^{n}\times \R^{n} \to \R_{\geqslant0} $, $\mathcal{K}_{\infty}$ functions $\underline{\alpha}, \overline{\alpha},\rho_{\omega_{c}},\rho_{\omega_{d}}, \rho_{u_{c}},\rho_{u_{d}}$, and constants $\kappa_{c}\in\R,\kappa_{d}\in\R$, such that the following hold,
\begin{itemize}
\item $\forall x,\hat x\in \R^{n}$,
\begin{align}\label{c1}
\underline{\alpha} (\| x-\hat{x}\| ) \leqslant V(x,\hat{x})\leqslant \overline{\alpha} (\| x-\hat{x}\|);
\end{align}
\item $\forall x,\hat x\in \R^{n}~\text{a}.\text{e}$, $\forall \omega,\hat{\omega}\in W$, and $\forall u,\hat u\in \mathbb{U}$,
\begin{align}\label{c2}
&\!\!\!\!\!\!\dfrac{\partial V(x,\hat{x})}{\partial x} \!f(x,\omega,u)\!+\!\dfrac{\partial V(x,\hat{x})}{\partial \hat{x}} \!f(\hat x,\hat{\omega},\hat{u})\\\notag
\!&\leqslant\! -\kappa_{c} V(x,\hat{x})\!+\!\rho_{\omega_{c}}(\| w\!- \!\hat{\omega}\| )\!+\!\rho_{u_{c}}(\| u\!-\! \hat{u}\|);
\end{align}\!\!\!\!\!
\item $\forall x,\hat x\in \R^{n}$,$\forall \omega,\hat{\omega}\in W$, and $\forall u,\hat u\in \mathbb{U}$,
\begin{align}\label{c3}
&\!\!\!\!\!V(g(x,\omega,u),g(\hat x,\hat{\omega},\hat{u}))\\\notag
\!&\leqslant\! \kappa_{d} V(x,\hat{x})+\rho_{\omega_{d}}(\| \omega\! -\! \hat{\omega}\| )+\rho_{u_{d}}(\| u \! -\! \hat{u}\| ).
\end{align}
\end{itemize}  
\end{assumption}
\begin{assumption}\label{ass2} 
There exist $\mathcal{K}_{\infty}$ function $\hat{\gamma}$ such that for all $x,y,z \in \R^{n}$,
\begin{align}\label{tinq} 
V(x,y)\leqslant V(x,z)+\hat{\gamma}(\| y-z\|).
\end{align}
\end{assumption}

We now have all the ingredients to construct a symbolic model $\hat{T}_{\tau}(\Sigma)$ of transition system $T_{\tau}(\Sigma)$ associated with the impulsive system $\Sigma$ admitting a function $V$ that satisfies Assumption \ref{likeiss} as follows. 
\begin{definition}\label{smm} Consider a transition system $T_{\tau}(\Sigma)=(X,X_{0}, W,\mathcal{W},U,\mathcal{U},\TF,Y,\op)$, associated to the impulsive system $\Sigma=(\R^{n},W,\mathsf{W},\mathbb U,\mathsf{U}_{\tau},f,g,\mathbb Y,h,\Omega)$. Assume $\Sigma$ admits a function $V$ that satisfies Assumption \ref{likeiss}. One can construct symbolic model $\hat T_{\tau}(\Sigma)=(\hat{X},\hat{X}_{0},\hat{W},\hat{\mathcal W},\hat{U},\hat{\mathcal U},\hat{\TF},\hat{Y},\hat{\op})$ where:
\begin{itemize}
\item  $\hat X=\hat{\R}^{n} \times \{0,\ldots,\overline{z}\}$, where $\hat{\R}^{n}=[\R^{n}]_{\eta^x}$ and $\eta^x$ is the state set quantization parameter;
\item  $\hat X_{0}=\hat{X}\times \{0\}$; 
\item $\hat{W}=[W]_{\eta^\omega}$, where $\eta^\omega$ is the internal input set quantization parameter;
\item $\hat{\mathcal W}=\{\hat{\omega}:[0,\tau]\rightarrow \hat{W}| \hat{\omega}\in{\mathcal W}_{\tau}|_{[0,\tau]}\}$;
\item $\hat{U}=[U]_{\eta^u}$, where $\eta^u$ is the external input set quantization parameter;
\item $\hat{\mathcal U}=\{\hat{u}:[0,\tau]\rightarrow \hat{U}| \hat{u}\in{\mathcal U}|_{[0,\tau]}\}$;
%			\item $\hat{\omega}=[W]_{\eta^w}$, where $0<\eta^w\leqslant\emph{span}(W)$ is the internal input set quantization parameter;
\item  $(\hat x^+,c^+)\in \hat\TF((\hat x,c),\hat{\omega},\hat u)$ iff one of the following scenarios hold:
\begin{itemize} 
\item Flow scenario: $0\leqslant c\leqslant \overline{z}-1$, $|\hat x^+-\mathbf{x}_{\hat{x},\hat{\omega},\hat{\nu}}(\tau)|\leqslant\eta^x$, and $c^+=c+1$;
\item Jump scenario: $\underline{z}\leqslant c\leqslant \overline{z}$,  $|\hat x^+ - g(\hat x,\hat{\omega}(0),\hat u(0))|\leqslant\eta^x$, and $c^+=0$; 
\end{itemize}
\item $\hat Y=Y$;
%\item $\hat\op:\hat X\rightarrow \hat Y$, defined as $\hat\op=\op$.
\item $\hat\op=\op$.
\end{itemize}
%			\begin{itemize}
%				\item $\hat{X}=\hat{X}\times \{0,\cdots,\overline{z}\}$, where $\hat{X}=[X]_{\eta}$ and $0<\eta\leqslant\emph{span}(X)$ is the state set quantization parameter; 
%				\item $\hat{X}_0=\hat{X}\times \{0\}$;
%				\item $\hat{U}=[U]_{\mu}$, where $0<\mu\leqslant\emph{span}(U)$ is the input set quantization parameter;
%				\item $(\hat{x}^+,c^+)\in \hat{\TF}((\hat{x},l),\hat{u})$ if and only if one of the following scenarios hold:
%				\begin{itemize} 
%					\item Flow scenario: $0\leqslant l\leqslant \overline{z}-1$, $|\hat{x}^+- \mathbf x(\tau,\hat{x},\hat{u})|\leqslant\eta$, and $l^+=l+1$;
%					\item Jump scenario: $\underline{z}\leqslant l\leqslant \overline{z}$, $|\hat{x}^+- g(\hat x,\hat{u})|\leqslant\eta$, and $c^+=0$;
%				\end{itemize}
%				\item $\hat{Y}=Y$;
%				\item $\hat{\op}:\hat{X}\rightarrow \hat{Y}$, defined as $\hat{\op}(\hat x,l)=\hat x$.
%			\end{itemize} 
\end{definition}
%An illustration of the computation of the transitions of $\hat T_{\tau}(\Sigma)$ is shown in Figure \ref{trans}. 
In the definition
of the transition function, and in the remainder of the paper, we abuse
notation by identifying $\hat u$ (respectively $\hat{\omega}$) with the constant external (respectively internal) input curve with domain
$[0,\tau)$ and value $\hat u$ (respectively $\hat{\omega}$).
%\begin{figure}[h]
%	\centering
%	\includegraphics[height=6cm, width=9cm]{allp}
%	\caption{Trajectories of system $T_{\mathsf{U}_\tau}(\Sigma)$ for different values of $p$, $a$, $b$, $d$, $\Psi$: \textcolor{blue}{blue} ($p=5$, $a=0.5$, $b=0.75$, $d=10$, $\Psi=\{25,50\}$), \textcolor{red}{red} ($p=10$, $a=-0.25$, $b=1.2$, $d=15$, $\Psi=\{50,75\}$), \textcolor{myco}{yellow} ($p=15$, $a=0.1$, $b=0.9$, $d=20$, $\Psi=\{75,100\}$)}
%	\label{allp}
%\end{figure}
%	\begin{figure}[t]
%		%\begin{framed} 
%		\centering
%		\subfigure[]{\includegraphics[height=3.5cm, width=4.cm]{p01}}
%		\subfigure[]{\includegraphics[height=3.5cm, width=4.3cm]{allp1}}
%		\subfigure[]{\includegraphics[height=3.5cm, width=4.3cm]{pp1}}
%		\vspace{-0.3cm}
%		%\end{framed}
%		\caption{{\small An illustration of the computation of the transitions of $\hat T_{\tau}({\Sigma})$ for particular $\hat x$ and $\hat u$ with (a) $l<\underline{z}$, (b) $\underline{z}\leqslant l\leqslant \overline{z}-1$, and (c) $l=\overline{z}$.}}
%		\label{trans}
%	\end{figure}
Now, we establish the relation from $T_{\tau}(\Sigma)$ to $\hat T_{\tau}(\Sigma)$, introduced above, via the notion of alternating simulation function as in Definition \ref{sf}.
\begin{theorem}\label{thm1}
Consider an impulsive system $\Sigma=(\R^{n},W,\mathsf{W},\mathbb U,\mathsf{U},f,g,\mathbb Y,h,\Omega)$ with its associated transition system $T_{\tau}(\Sigma)=(X,X_{0}, W,\mathcal{W},U,\mathcal{U},\TF,Y,\op)$. Suppose Assumptions \ref{likeiss}, \ref{ass2}, and \ref{h} hold. Consider symbolic model $\hat T_{\tau}(\Sigma)=(\hat{X},\hat{X}_{0},\hat{\omega},\hat{\mathcal W},\hat{U},\hat{\mathcal U},\hat{\TF},\hat{Y},\hat{\op})$ constructed as in Definition \ref{smm}. If inequality,
\begin{align}\label{cc1}
\ln(\kappa_{d})-\kappa_{c}\tau c<0,
\end{align}
holds for $c\in\{\underline{z},\overline{z}\}$, then function $\mathcal{V}$ defined as,
\begin{align}\label{sm}
\mathcal{V}((x,c),(\hat{x},c))\!\!\Let\!\!\left\{
\begin{array}{lr} 
\!\!\!\!V(x,\hat{x})    \quad\,~ \text{if}\quad \kappa_{d}<1 \!~\&~ \kappa_{c}>0,\\
\!\!\!\!\dfrac{V(x,\hat{x})}{e^{-\kappa_{c}\tau \epsilon c}}   \quad \text{if}\quad \kappa_{d}\geqslant1 ~\&~ \kappa_{c}>0,\\
\!\!\!\!\dfrac{V(x,\hat{x})}{\kappa_{d}^{-\frac{c}{\delta}}} \quad\, \text{if}\quad \kappa_{d}<1 ~\&~ \kappa_{c}\leqslant0,
\end{array}\right.
\end{align}
for some $0<\epsilon<1$ and $\delta>\overline{z}$, is an alternating simulation function from $\hat T_{\tau}(\Sigma)$ to $T_{\tau}(\Sigma)$. 
\end{theorem}

\section{Case study}\label{sec:Examples}

Consider the exchange problems between $N$ interconnected warehouses of a storage-delivery process. Denote by $\textbf{x}_{i} \in \mathbb{R}_{\geq 0}$, the number of goods in the warehouse $ i $. The interconnections between the warehouses is supposed to be circular.

\underline{Under the flow mode:} When $ t\in \mathbb{R}_{\geqslant 0 }\backslash \Omega_i $, for each warehouse the state $ x_{i} $ is continuously controlled through a delivery and picking-up process with a quantity $ d_{i} $ and input signal $\nu_{i}(t) \in \{-1,1\}, t \in[0, \tau)$.
% \begin{itemize}
% \item[$\bullet$] For warehouse 1, the state variable $x_{1}$ evolves continuously, proportional to the number of goods it contains, at a rate given by the coefficient $a_{1}$. Additionally, the state of warehouse 3 also impacts the evolution of $ x_{1} $, through a continuous provision with rate coefficients $ b_{1} $. 
% \item[$\bullet$] For warehouse 2, the state variable $x_{2}$ evolves continuously, proportional to the number of goods it contains, at a rate given by the coefficient $a_{2}$. Additionally, the state of warehouse 1 also impacts the evolution of $ x_{2} $, through a continuous provision with rate coefficients $ b_{2} $. 
% \item[$\bullet$] For warehouse 3, the state variable $x_{3}$ evolves continuously, proportional to the number of goods it contains, at a rate given by the coefficient $a_{3}$. Additionally, the state of warehouse 2 also impacts the evolution of $ x_{3} $, through a continuous provision with rate coefficients $ b_{3} $.   
% \end{itemize}

\underline{Under the jump mode:} At each time $t \in \Omega_{i}=\left\lbrace t_k^{i}\right\rbrace_{k \in \mathbb{N}, i=1,2,3}$, with $t_{k+1}^{i}-t_{k}^{i} \in\left\lbrace\underline{z}_{i} \tau_{i}, \ldots, \bar{z}_{i} \tau_{i}\right\rbrace$ for fixed jump parameters $\tau_{i} \in \mathbb{R}_{> 0}$ and $\underline{z}_{i}, \bar{z}_{i} \in \mathbb{N}_{\geq 1}, \underline{z}_{i} \leq \bar{z}_{i}, $, a truck enters warehouse $ i $ and the state $ x_{i} $ becomes controlled through a delivery and picking-up process with a quantity $ \bar{d}_{i} $ and input signal $\nu_{i}(t) \in \{-1,1\}, t \in[0, \tau)$. 

% Additionally, we consider an exchange scenario among the three warehouses as follows:
% \begin{itemize}
% \item[$\bullet$] For warehouse 1, the state variable $x_{1}$ evolves proportional to the number of goods it contains, at a rate given by the coefficient $r_{1}$. Additionally, the state of warehouse 3 also impacts the evolution of $ x_{1} $, through a continuous provision with rate coefficients $ q_{1} $. 
% \item[$\bullet$] For warehouse 2, the state variable $x_{2}$ evolves continuously, proportional to the number of goods it contains, at a rate given by the coefficient $r_{2}$. Additionally, the state of warehouse 1 also impacts the evolution of $ x_{2} $, through a continuous provision with rate coefficients $ q_{2} $. 
% \item[$\bullet$] For warehouse 3, the state variable $x_{3}$ evolves continuously, proportional to the number of goods it contains, at a rate given by the coefficient $r_{3}$. Additionally, the state of warehouse 2 also impacts the evolution of $ x_{3} $, through a continuous provision with rate coefficients $ q_{3} $.  
% \end{itemize}
The full state of each warehouse $ x_{i} $ is observable and we assume that the interconnected system is realisable. The dynamic motion of this process in the case $N=3$ is modeled by, 

\begin{align*}
&\Sigma_{i}:\begin{cases}
\dot{\mathbf{x}}_{i}(t) =a_{i} \mathbf{x}_{i}(t)+b_{i}\mathbf{x}_{\Bar{i}}(t)+d_{i} \nu_{i}(t), & t \in \mathbb{R}_{\geq 0} \backslash \Omega_{i}, \\
\mathbf{x}_{i}(t)=r_{i} \mathbf{x}_{i}(t^{-})+q_{i} \mathbf{x}_{\Bar{i}}(t) +\bar{d}_{i} \nu_{i}(t), & t \in \Omega_{i} ,\\
\mathbf{y}_{i}(t)=\mathbf{x}_{i}(t).& 
\end{cases}
\end{align*}
with $ i=1,\dots,N$ and $\Bar{i}=\begin{cases} i-1 & i>1\\
N &  i=1
\end{cases}$.
In order to construct a symbolic model for the interconnected impulsive systems, we have to check Assumptions \ref{sg}, \ref{likeiss}, \ref{ass2} and \ref{h}. 

In the sequel, we will only detail the shell for the case $N=3$. It can be shown that conditions \eqref{c1}, \eqref{c2} and \eqref{c3} hold for each subsystem $ \Sigma_{i} $ with $ V_{i}\left(x_{i}, x_{i}^{\prime}\right)=\left\|x_{i}-x_{i}^{\prime}\right\|, \; i=1,2,3$, with,   $\underline{\alpha}_{i}=\bar{\alpha}_{i}=\mathcal{I}_d, \kappa_{c_{i}}=-a_{i}, \kappa_{d_{i}}=|r_{i}| $, $\rho_{u_c, 1}=|d_{1}|$, $\rho_{u_d,1}=|\bar{d}_{1}| $, $ \rho_{\omega_{c},1}=|b_{1}| $, $ \rho_{\omega_{d},1}=|q_{1}| $, $\rho_{u_c, 2}=|d_{2}|$, $\rho_{u_d,2}=|\bar{d}_{2}| $, $ \rho_{\omega_{c},2}=|b_{2}| $, $ \rho_{\omega_{q},2}=|q_{2}| $,  $\rho_{u_c, 3}=|d_{3}|$, $\rho_{u_d,3}=|\bar{d}_{3}| $, $ \rho_{\omega_{c},3}=|b_{3}| $ and $ \rho_{\omega_{d},3}=|q_{3}| $. From these functions, we can drive the expressions of the $ \gamma_{ij} $ functions in Assumption \ref{sg}. Thus, $ \gamma_{31}=\max\left\lbrace |b_{1}|,|q_{1}| \right\rbrace $, $ \gamma_{12}= \max\left\lbrace |b_{2}|,|q_{2}| \right\rbrace $ and $ \gamma_{23}= \max\left\lbrace |b_{3}|,|q_{3}| \right\rbrace $.

% \begin{remark}
% To verify Assumption \ref{sg}, we can project the bounded-real lemmas for both continuous-time and discrete-time systems, as described in \cite{boyd1994linear}, in the form of convex linear matrix inequality on each subsystem $ \Sigma_{i} $. 
% \begin{itemize}
% \item[$\bullet$] For the flow mode, we have $\left\|\Sigma_i^{t \in \mathbb{R}_{\geqslant 0} \backslash \Omega_{i}}\right\|_{\infty}<\gamma_{ij}^{F}$ iff there exists $P_{i} \geqslant 0$ such that,
% \begin{align}\label{FlowLMI}
% \left[\begin{array}{cc}
% a_{i}^\TT P_{i}+P_{i} a_{i} +I & P_{i} b_{i}   \\
% b_{i}^\TT P_{i}           & -(\gamma_{ij}^{F})^{2} I  
% \end{array}\right]  \preccurlyeq 0;
% \end{align}
% \item[$\bullet$] For the jump mode, the constraint $\left\|\Sigma_i^{t \in \Omega_{i}}\right\|_{\infty}<\gamma_{i}^{J}$ is equivalent to the existence of $P_{i}>0$ such that,
% \begin{align}\label{JumpLMI}
% \left[\begin{array}{cc}
% r_{i}^\TT P_{i} r_{i}-P_{i}+I & r_{i}^\TT P_{i} q_{i} \\
% q_{i}^\TT P_{i} \bar{A} & q_{i}^\TT P_{i} q_{i}-(\gamma_{ij}^{J})^{2} I
% \end{array}\right] \preccurlyeq 0.
% \end{align}
% Note that $\gamma_{ij}$ for each $ \Sigma_{i} $ is defined by $ \gamma_{ij}=\max\left\lbrace \gamma_{ij}^{F}, \gamma_{ij}^{J}\right\rbrace  $.\hfill $\diamond$
% \end{itemize} 
% \end{remark}

Assumption \ref{ass2} holds with $\hat{\gamma}=\mathcal{I}_d$ and Assumption \ref{h}, is satisfied with $ L=1 $. Now, given $ \tau_{i} $ and $ c_{i} $ satisfying \eqref{cc1} for $c_{i} \in\{\underline{z}_{i}, \bar{z}_{i}\}$, and, with a proper choices of $\epsilon_{i}$ and $\delta_{i}$, functions $\mathcal{V}_{i}(x_{i}, \hat{x}_{i})$ given by \eqref{sm} are local alternating simulation functions from $\hat{T}_\tau(\Sigma_{i})$, constructed as in Definition \ref{smm} for each $ i^{th} $ subsystem $ i=1,2,3 $, to $T_\tau(\Sigma_{i})$. In particular, each $\mathcal{V}_{i}$ satisfies conditions \eqref{lsf1} and \eqref{lsf2} with functions $\alpha_{i}, \; \bar{\rho}_{\omega_{i}}, \bar{\rho}_{u_{i}},$ and constants $\bar{\sigma}_{i}, \varepsilon_{i}$ given below based on the values of $ a_{i}$ and $r_{i}$, with $\psi=0.99$.
\begin{itemize}
\item  $|r_{i}|<1 \:\&\: a_{i}<0: \alpha_{i}=\mathcal{I}_d,  \tilde{\sigma}_{i}=\max \left\lbrace e^{a_{i}\tau_{i}}, r_{i}\right\rbrace, \bar{\rho}_{\omega_{i}}=\max \left\lbrace  b_{i}, q_{i} \right\rbrace, \rho_{u_{i}}=0,  \varepsilon_{i}=\hat{\varphi}_{i}$.
\item  $|r_{i}|\geqslant1 \:\&\: a_{i}<0: \alpha_{i}=\mathcal{I}_d, \rho_{u_{i}}=\rho_{\omega_{i}}=0, \bar{\sigma}_{i}=\max \left\lbrace e^{a_{i}\tau_{i}(1+\epsilon_{i}c_{i})}, e^{a_{i}\tau_{i}\epsilon_{i}c_{i}}|r_{i}|\right\rbrace, \varepsilon_{i}=\!e^{\kappa_{c}\tau \epsilon (\overline{z}+1)}\!\hat{\varphi}$.
\item $|r_{i}|<1 \:\&\: a_{i}\geqslant0: \alpha_{i}=\mathcal{I}_d, \rho_{u_{i}}=\rho_{\omega_{i}}=0, \bar{\sigma}_{i}=\max \left\lbrace e^{a_{i}\tau_{i}}|r_{i}|^{\frac{c_{i}}{\delta_{i}}}, |r_{i}|^{\frac{\delta_{i}+c_{i}}{\delta_{i}}}\right\rbrace, \varepsilon_{i}= \hat{\varphi}_i$.
\end{itemize}

\begin{figure}
\small
    \centering
    \includegraphics{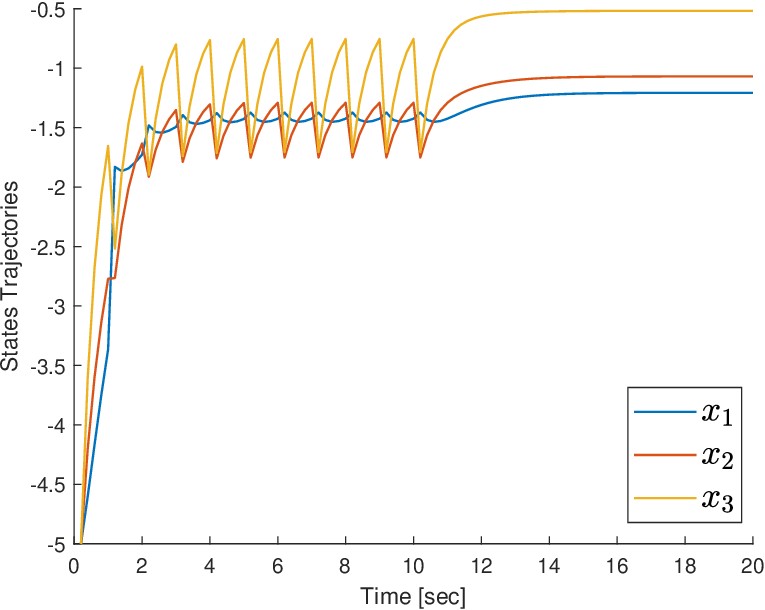}
    \caption{State trajectories under fixed point controller.}
    \label{fig:statetrajectories}
\end{figure}

The control objective is to maintain the number of items of each warehouse $ i $ in a desired range $O_{i}$ given by $O_{i}=\left[\ominus_{min}, \ominus_{min}\right]$ (a safety specification). We set up the system with the following parameters $a1 = -1, \;
b_{1} = 0.4, \;
d_{1} = 1, \;
r_{1} = 0.05, \;
q_{1} = 0.4, \;
\bar{d}_{1} = 1, \;
a_{2} = -1.5, \;
b_{2} = 0.5, \;
d_{2} = 1, \;
r_{2} = 0.03, \;
q_{2} = 0.5, \;
\bar{d}_{2} = 1, \;
a_{3} = -2, \;
b_{3} = 0.5, \;
d_{3} = 0.5, \;
r_{3} = 0.08, \;
q_{3} = 0.5, \;
\bar{d}_{3} = 1, \;$ and consider the following, for $i=1,\dots,3$,  
$
\Omega_{i}= \begin{bmatrix}
	1 & 2 & 3 & 4 & 5 & 6 & 7 & 8 & 9 & 10
\end{bmatrix};$ Each system state is expected to operate around an equilibrium point within the range of $\begin{bmatrix}
	-5 & 5
\end{bmatrix}$. With the defined system parameters, the sampling period for the controller to be designed is set $\tau = 0.2$, which satisfies condition \eqref{cc1} for all the subsystems. We discretize the state by $n^x=0.6667$. 
We conducted both monolithic and compositional abstractions, with the former taking $3589$ seconds and the latter taking $1546$ seconds to compute. Figure \ref{fig:statetrajectories} displays the state trajectories using the designed fixed-point controller \cite{cassandras2009introduction}. It is evident from the figure that the designed controller successfully keeps the states within the required safe region.

We compared computation times between monolithic and compositional abstractions for varying subsystem numbers (Table \ref{tab:ComparisonComputationalTime}). Results show computation times in seconds for each abstraction and subsystem count, at a discretization parameter $n^x=2.5$. Compositional abstraction generally requires less time than monolithic, even as subsystems increase. The time difference remains significant; for instance, with five subsystems, compositional abstraction is almost six times faster. This makes it more computationally efficient, particularly when dealing with numerous subsystems.

\begin{table}
		\caption{Abstraction Computation Time Comparison [s]}
		\centering
%	\setlength\tabcolsep{3.5pt}\renewcommand\arraystretch{1.25}
%	\noindent\makebox[\textwidth]{%
		\begin{tabular}{|p{2.5cm}|*{4}{c|}}
			\hline
			\diagbox[width=3cm, height=0.5cm]{\tiny Abstraction}{ \tiny Number of subsys. } &  2   & 3 & 4 & 5 \\ \hline
			Monolithic                                                                                                           & $ 0.3107 $ &    $ 1.2285 $     &  $ 13.0902 $ & $ 5453.65 $      \\ \hline
		Compositional &    $ 0.2108 $     &   $ 0.3147 $      &    $ 2.2348 $ & $ 975.4288$ 
        \\ \hline
		ratio &    $ 1.4739 $     &   $ 3.9037 $      &    $ 5.8574 $ & $ 5.5910$  \\ \hline
		\end{tabular}
	%}
\label{tab:ComparisonComputationalTime}
\end{table}

% This time difference can be attributed to the fact that the compositional approach eliminates overall states when there is no successor for a sub-state, which leads to a more compact representation. Additionally, the corresponding sparse matrix for the compositional abstraction was smaller in size at 8296567 bytes compared to the monolithic abstraction's matrix of 11785057 bytes, which highlights the advantages of using the compositional approach in terms of memory usage. 

\section{Conclusion}
To conclude, this paper introduces a novel compositional technique for building symbolic models in interconnected impulsive systems using the concept of approximate alternating simulation function. With certain small gain-type conditions, our method compositionally establishes an overall alternating simulation function, connecting interconnection symbolic models and original impulsive subsystems. Moreover, we present a method, guided by stability and forward completeness, to create symbolic models with corresponding alternating simulation functions for impulsive subsystems.

Future work involves extending this approach to stochastic impulsive systems, integrating probabilistic distributions for characterizing flow and jump mode functions.

%\vspace{-5pt}
\bibliographystyle{ieeetr}      % Include this if you use bibtex 
\bibliography{arxive}

\end{document}